\documentclass[noblind]{geophysics}
\usepackage{mathrsfs}
\usepackage[linesnumbered,ruled,vlined]{algorithm2e}
\usepackage{amsmath}
\usepackage{amssymb}
\usepackage{booktabs}
\usepackage[colorlinks,linkcolor=black,anchorcolor=black,citecolor=black]{hyperref}

\begin{document}

\setfigdir{.}

\newcommand{\rs}[1]{\mathstrut\mbox{\scriptsize\rm #1}}
\newcommand{\rr}[1]{\mbox{\rm #1}}

\title{Iterative separation of coherent blended signals in common shot gathers using synchrosqueezed curvelet-Radon constraints}

\renewcommand{\thefootnote}{\fnsymbol{footnote}}
\author{
Zifei Li\footnotemark[1]\footnotemark[3]
,
Shaohuan Zu\footnotemark[1]\footnotemark[2]\footnotemark[3]
, and 
Haojun Chen\footnotemark[1]\footnotemark[3]
}

\footnotetext[1]{State Key Laboratory of Oil and Gas Reservoir Geology and Exploitation, Chengdu University of Technology, Chengdu, 610059, China}
\footnotetext[2]{Key Laboratory of Earth Exploration and Information Technology of Ministry of Education, Chengdu University of Technology, Chengdu, 610059, China}
\footnotetext[3]{College of Geophysics, Chengdu University of Technology, Chengdu, 610059, China}

\address{
\footnotemark[1]State Key Laboratory of Oil and Gas Reservoir Geology and Exploitation, Chengdu University of Technology,
Chengdu, 610059, China \\
\footnotemark[2]Key Laboratory of Earth Exploration and Information Technology of Ministry of Education,
Chengdu University of Technology,
Chengdu, 610059, China \\
\footnotemark[3]College of Geophysics,
Chengdu University of Technology,
Chengdu, 610059, China 
	}

\righthead{CSG Iterative Deblending }

\begin{abstract}
    Blended data acquired via simultaneous-source seismic exploration conventionally require post-acquisition deblending, which typically relies on the coherence differences introduced by firing time delays (time dithering). To reduce the dependency of the deblending process on these time dithers, a novel joint constraint based on the synchrosqueezed transform and the Radon transform is proposed, operating directly in the common-shot gather (CSG) domain. Specifically, by exploiting the differences in propagation directions of the blended signals within CSGs, the synchrosqueezed transform is first employed for an initial iterative separation to extract individual sources directly from the continuous records. Once the majority of the valid signals are separated, the Radon transform is subsequently applied in further iterations to suppress the residual blending interference, thereby preventing amplitude damage to the effective signals. Compared to conventional non-CSG deblending methods, this approach bypasses the reliance on time-delay coherence differences, thus enabling real-time quality monitoring of individual sources during field acquisition. Furthermore, compared to other existing CSG-based separation techniques, the proposed joint-constraint iterative framework demonstrates superior performance when handling complex data. Applications on both synthetic and field blended datasets demonstrate that high-fidelity data separation can be successfully achieved independently of the time-dithering constraints. Finally, because the proposed method operates completely independently across different CSG slices, it is highly amenable to parallel computing, facilitating the efficient processing of massive datasets within a short timeframe.

\end{abstract}

\section{Introduction}
The demand for enhanced operational efficiency and reduced acquisition costs has propelled simultaneous-source technology to the forefront of modern seismic exploration \cite[]{bagainiOverviewSimultaneousVibroseis2006}. By allowing multiple sources to be activated with overlapping recording times, this acquisition paradigm significantly increases spatial sampling density and reduces survey duration\cite[]{mooreSimultaneousSourceSeparation2008}. Despite its profound economic and imaging advantages, the simultaneous firing of sources fundamentally introduces severe overlapping interference---commonly referred to as blending noise or crosstalk---among consecutive excitations\cite[]{DaveHoweIndependent}. Consequently, decoupling these composite continuous wavefields to preserve high-fidelity structural information remains a critical and challenging prerequisite in the standard seismic processing pipeline.

Regardless of whether direct filtering or inversion-driven frameworks are employed, the efficacy of wavefield separation fundamentally relies on the kinematic incoherency induced by pseudo-random time-dithering schemes \cite[]{abmaSimultaneousSourceSeismic2020}. In cross-sorted gathers, such as common-receiver (CRG), common-offset (COG), common-midpoint (CMP), and common-channel gathers (CCG), this encoded temporal irregularity translates into distinct coherence discrepancies between the primary reflections and the overlapping interference\cite[]{abmaOverviewBPMarine2012}. Extensive research has exploited these dither-induced characteristics to optimize crosstalk attenuation. For instance, filtering strategies have been tailored to capitalize on strong signal coherency in the CMP domain \cite[]{zhangDeblendingUsingHighresolution2015, wang2023deblending} or augmented with structural dip constraints in the CRG domain \cite[]{yangDeblendingWeakSignal2017}. Concurrently, inversion-based methodologies formulate separation as an iterative predictive subtraction problem \cite[]{mahdad2011, mahdadIterativeMethodSeparation2012}, often stabilized by advanced mathematical solvers such as shape regularization \cite[]{chenIterativeDeblendingMultiple2015} or fast iterative shrinkage-thresholding \cite[]{qu2016}.

Beyond conventional mathematical transformations, the distinct interference patterns generated by time dithering can be effectively mapped using deep learning architectures. By embedding neural networks as data-driven priors within the iterative deblending framework, standard sparse constraints can be substituted with learned feature extractions \cite[]{zu2020}. To further enhance a network's capacity to discriminate these dither-induced discrepancies, \cite{wangIterativeDeblendingUsing2022} introduced a data augmentation protocol utilizing shuffled crosstalk and approximate labels. Additionally, sophisticated network topologies, such as the Multi-Res UNet, have been deployed in the CRG domain to execute joint iterative separation and missing trace reconstruction \cite[]{wangDeblendingRecoveryIncomplete2022,wangDeblendingSeismicData2022,wangSelfSupervisedDeepLearning2023}. Comprehensive multi-data training strategies, low-amplitude interference, and unblended targets, have also been developed to fortify prediction stability and separation accuracy \cite[]{wangMultidataTrainingMethod2023}.

Despite these algorithmic advances, the practical performance of incoherency-based separation is strictly bottlenecked by the physical limits of the time-dithering sequence during acquisition. When the temporal delays lack sufficient irregularity, the required coherence discrepancy diminishes, inevitably degrading the separation fidelity \cite[]{zuInfluenceSimultaneoussourceSurvey2022}. Because these methodologies explicitly depend on dither-induced incoherency, they must be executed in cross-sorted domains (e.g., CRG or COG), which inherently prohibits direct separation within the common-source gather (CSG) domain where the interference remains highly coherent. Alternatively, deep learning-based methods have also been proposed for separation \cite{sunjingdeblendingcsg}. However, whether supervised or self-supervised, these approaches inherently rely on massive volumes of acquired seismic data. This limitation makes real-time quality control (QC) of individual source excitations during active field operations practically impossible. Although cross-domain iterative strategies \cite{muMixedDomainDeblending2024} can partially mitigate resolution loss, they fundamentally depend on dither-derived features and thus remain vulnerable to inadequate temporal staggering. Alternative source-encoding paradigms, such as the popcorn shooting technique \cite[]{abmaPopcornShootingSparse2013} and periodically varying codes \cite[]{zuPeriodicallyVaryingCode2016}, have proven effective at enhancing interference irregularity without relying solely on time dithering. However, these methodologies mandate complex modifications to standard field acquisition protocols.

To circumvent the strict dependency on randomized time delays without requiring specialized acquisition hardware, this paper introduces a novel separation framework that operates directly on the unseparated continuous records within the CSG domain. By incorporating joint constraints derived from the synchrosqueezing and Radon transforms, the proposed methodology utilizes an iterative inversion scheme to explicitly decouple the coherent blending interference. This direct CSG-domain processing not only suppresses noise residues in continuously coherent regions but also enables the on-the-fly monitoring of isolated single-source quality during active acquisition. Evaluations performed on both synthetic models and field-acquired blended datasets demonstrate that the proposed workflow achieves high-fidelity crosstalk attenuation entirely independent of the randomized time-delay parameters. Additionally, because the algorithmic operations are completely independent across individual CSG volumes, the framework allows for massive parallelization, ensuring rapid computational turnaround for large-scale seismic surveys.

\section*{Method}

\subsection{Iterative Deblending in CSG}
Compared to conventional acquisition, blended acquisition provides a compelling advantage in both economic cost and operational efficiency. In conventional surveys, individual sources are fired sequentially with sufficient time intervals to prevent wavefield overlapping. Conversely, blended acquisition allows for overlapping recording times, where the receivers $r_i$ ($i = 1, 2, \dots, N$) continuously record the superimposed wavefields generated by multiple simultaneous sources $s_j$ ($j \in S$). The continuously recorded blended data $\mathbf{d}(t, r_i)$ can be mathematically expressed as:
\begin{equation}
    \mathbf{d}(t, r_i) = \sum_{j \in S} \mathbf{d}_j(t - \tau_j, r_i),
\end{equation}
where $\mathbf{d}_j$ represents the unblended true signal associated with the $j$-th source, and $\tau_j$ denotes its corresponding random firing time delay (i.e., time dithering).

To formulate the subsequent separation algorithm, the physical superposition described above is generally abstracted into a compact matrix-operator form. Specifically, the unblended signals from all simultaneous sources are concatenated into a higher-dimensional matrix $\mathbf{m}$. The temporal shifting and blending operations are mathematically encapsulated within a global forward-blending operator $\mathbf{\Gamma}$, allowing the continuous blended data to be concisely denoted as $\mathbf{d} = \mathbf{\Gamma} \mathbf{m}$.

During the separation process, the wavefield corresponding to the target source is regarded as the effective signal, while the overlapping signals originating from all other simultaneous sources are treated as blending interference (crosstalk). In the common-shot gather (CSG) domain, this blending interference exhibits coherence characteristics identical to those of the effective signal, as the overlapping records merely experience vertical temporal shifts dictated by their firing times (Figs.~\ref{fig:fff1} and \ref{fig:fff2}). Because this strong coherence fundamentally hinders direct separation, conventional deblending algorithms are typically executed in cross-sorted domains (e.g., common-receiver gathers), where the random time-dithering schemes disrupt the coherence of the crosstalk. By applying the pseudo-deblending (adjoint) operator $\mathbf{\Gamma}^T$ to the continuous data, the initial pseudo-deblended data can be obtained as $\mathbf{d}_{\mathrm{pseudo}} = \mathbf{\Gamma}^T \mathbf{d}$. In these non-CSG domains, the conventional iterative separation framework typically updates the high-dimensional source model $\mathbf{m}$ using the iterative shrinkage-thresholding algorithm (ISTA), which can be formulated as:
\begin{equation}
    \mathbf{m}_{l+1} = \mathcal{S}\left( \mathbf{m}_l + \mathbf{\Gamma}^T \left( \mathbf{d} - \mathbf{\Gamma}\mathbf{m}_l \right) \right),
\end{equation}
where $\mathbf{m}_l$ denotes the estimated high-dimensional signal ensemble at the $l$-th iteration, and $\mathcal{S}$ represents the sparsity-promoting constraint operator.

\multiplot*{2}{fff1,fff2}{width=0.4\columnwidth}
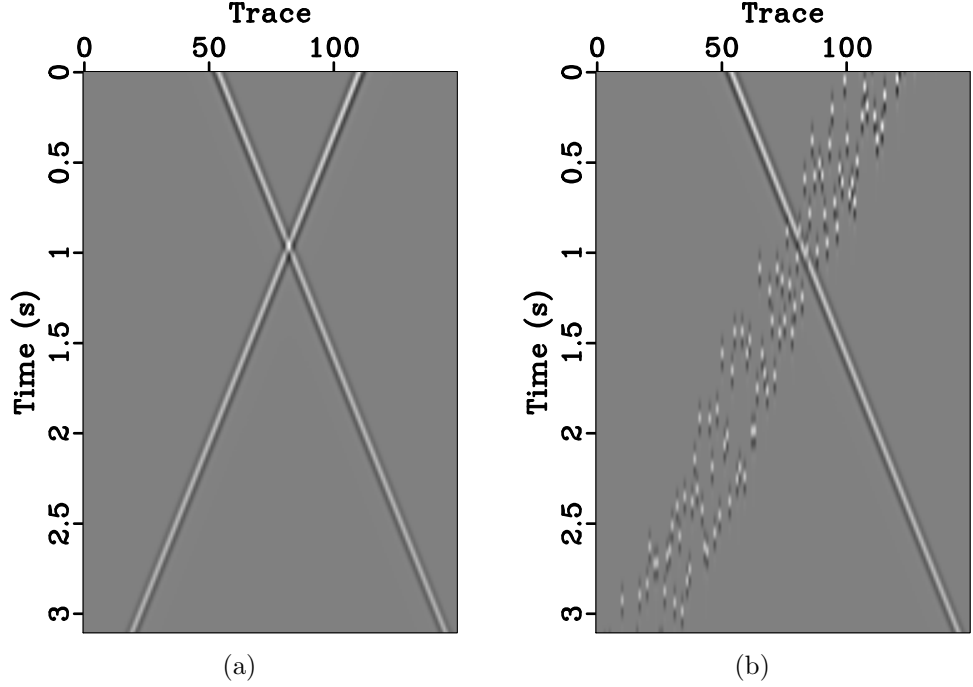
{Differences in blended data between CSG and non-CSG domains: (a) CSG; (b) non-CSG.}

Simply put, the core mechanism of the iterative framework lies in separating the noise from the effective signal as thoroughly as possible. This involves a trade-off: either allowing some residual noise to ensure the effective signal remains undamaged, or completely suppressing the noise at the cost of mildly damaging the effective signal. Because blending interference is transformed into incoherent noise via time dithering operators in the CRG domain, the iteration must be performed on pseudo-deblended records. The cleanly constrained signal must be forward-modeled (delayed) into interference noise and then correspondingly subtracted from the raw records.

However, in the CSG—or equivalently, the received continuous records—every individual source record remains continuously coherent and relatively complete. Consequently, the iterative framework can adopt a different logic: multiple clean source signals can be sorted out from a single continuous gather. When one source is selected as the primary target, all other signals can be directly subtracted from the original raw record to obtain the pseudo-separated record corresponding to that single source. This process can be expressed by the following equation:
\begin{equation}
    \mathbf{m}_{l+1} = \mathbf{\Gamma}^T\mathcal{S}\left(\mathbf{d} - \mathbf{\Gamma}\mathbf{m}_l\right)
\end{equation}
Here, $\mathbf{D}$ represents the raw continuous blended data, and $\mathbf{m}_l$ denotes the estimated interference model at the $l$-th iteration, which is essentially the summation of the estimated signals from all other non-target sources.

Within this iterative formulation, the constraint operator $\mathcal{S}$ no longer needs to exploit the coherence differences induced by $\mathbf{\Gamma}^T$. Instead, the role of $\mathbf{\Gamma}^T$ is strictly reduced to remapping the separated results back to their corresponding temporal positions within the original continuous records. In the proposed framework, $\mathcal{S}$ corresponds to the joint application of the 2-D synchrosqueezed curvelet transform (SSCT) clustering and the Radon transform constraints.

Within each iteration loop, the subtraction $(\mathbf{D} - \mathbf{m}_l)$ yields a progressively cleaner target record. First, the SSCT projects this record into the multi-dimensional phase space (MDPS), where clustering algorithms utilize localized spatial distances and instantaneous angular differences to robustly isolate the target source's coherent energy from the residual discontinuous noise. Subsequently, the Radon transform acts as a secondary filter in the intercept-slowness ($\tau-p$) domain, systematically scanning and attenuating any remaining unfocused interference to preserve the primary signal's true amplitude. By cyclically iterating this direct-subtraction and dual-constrained filtering process across all sources, the energy of the blending interference is monotonically reduced. The unblended signals are iteratively updated until the residual energy converges, ultimately yielding highly accurate, non-destructive reconstructions of the clean seismic records.

\subsection{Synchrosqueezed Curvelet Transform(SSCT)}
In blended common-shot gathers (CSGs), seismic records often exhibit intersecting signals with curved components in the space-time domain. As illustrated in the Fig~\ref{fig:fig2_wiggle} blending interference within the CSGs frequently manifests astruncated events with distinct propagation directions.
\plot*{fig2_wiggle}{width=0.8\columnwidth}
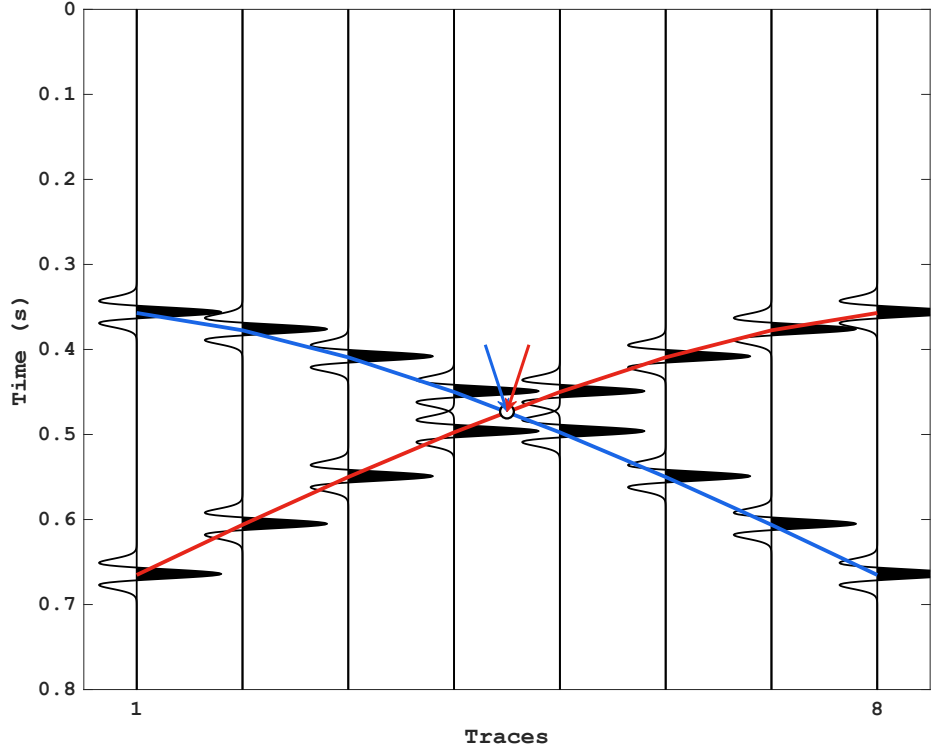
{Typical characgteristics of blended records in a common-shot gather: signals originating from a single source manifest as curved events, whereas interference signals directly truncate the coherent useful signals along their propagation directions. Although reflections of the main signal may intersect one another in areas with complex geological conditions, the blending interference remains significantly more severe and exhibits a broader spatial footprint.}

Utilizing the curvelet transform, these signals admit a sparse representation characterized by scale $\mathbf{j}$, angle $\mathbf{l}$, and spatial position $\mathbf{p}$. Similar to the Fourier transform, the curvelet transform essentially establishes a set of basis functions and calculates the correlation between the space-time domain signals and these bases. These correlation values constitute the representation of the signal within the curvelet frame. In the frequency-wavenumber domain $\boldsymbol{\omega} = (\omega_x,\omega_t)$, the basis functions are typically defined as:
\begin{equation}
    \hat{\varphi}_{\mathbf{j},\mathbf{l},\mathbf{p}}(\boldsymbol{\omega}) = \hat{\mathbf{U}}_\mathbf{j}(\mathbf{R}_{\theta_\mathbf{l}} \boldsymbol{\omega}) \cdot e^{-\mathrm{i} \langle \mathbf{x}_p^{(\mathbf{j},\mathbf{l})}, \boldsymbol{\omega} \rangle}
\end{equation}
Here, $\mathbf{R}_{\theta_\mathbf{l}}$ denotes the rotation matrix with an angle $\theta_\mathbf{l} = \mathbf{l} \cdot \frac{\pi}{2} \cdot 2^{-\lfloor \mathbf{j}/2 \rfloor}$, which controls the orientation of the curvelet. The term $\mathbf{x}_p^{(\mathbf{j},\mathbf{l})}$ represents the discrete translation vector for a given scale $\mathbf{j}$ and angle $\mathbf{l}$ at the current position $\mathbf{p}$ (Fig~\ref{fig:curvelet_wedge_tiling}). Furthermore, the wedge-shaped window $\hat{\mathbf{U}}_\mathbf{j}$ is composed of the product of the radial window $\mathbf{W}$ and the angular window $\mathbf{V}$:
\begin{equation}
    \hat{\mathbf{U}}_\mathbf{j}(r, \theta) = \alpha \mathbf{W}(r) \mathbf{V}(2^{\lfloor \mathbf{j}/2 \rfloor} \theta)
\end{equation}
where $r$ and $\theta$ denote the polar coordinate components in the $\boldsymbol{\omega}$ domain, and $\mathbf{W}$ and $\mathbf{V}$ represent the radial and angular windows constructed from Meyer wavelets, respectively. The parameter $\alpha$ acts as a scale-dependent normalization factor. According to Parseval's theorem, the inner product between the seismic signal $\hat{f}(\boldsymbol{\omega})$ and the complex conjugate of these basis functions yields the sparse representation of the seismic data in the curvelet domain:
\begin{equation}
    \mathbf{C}(\mathbf{j},\mathbf{l},\mathbf{p}) = \int \hat{f}(\boldsymbol{\omega}) \overline{\hat{\varphi}_{\mathbf{j},\mathbf{l},\mathbf{p}}(\boldsymbol{\omega})} \, d\boldsymbol{\omega}
\end{equation}

\plot*{curvelet_wedge_tiling}{width=0.8\columnwidth}
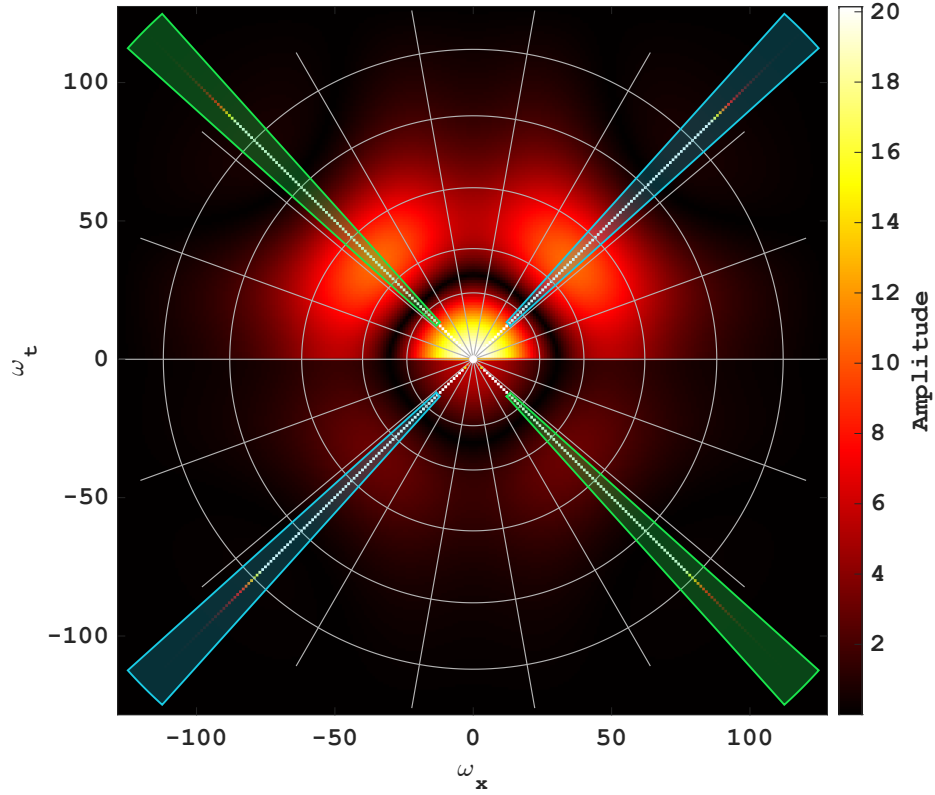
{Curvelet tiling and energy distribution in the frequency-wavenumber domain. The background displays the amplitude spectrum of the blended seismic record. The white grid illustrates the discrete curvelet tiling, where the radial window $\mathbf{W}$ and the angular window $\mathbf{V}$ partition the $\boldsymbol{\omega}$ domain into multiple scales $\mathbf{j}$ and angle $\mathbf{l}$. Through this basis function partitioning, the spectral peaks (bright lines) within the wedge-shaped regions can effectively isolate intersecting seismic events into distinct directional sectors.}

In blended CSGs, seismic records often contain signals from multiple sources that overlap at the same locations within the $x-t$ domain. However, these signals typically exhibit distinct dip angles, which manifest as varying rates of change in their curvelet coefficients with respect to spatial positions. Fig.~\ref{fig:scoef_at_point} plots the curvelet coefficients at various spatial positions extracted from the 10 $th$ scale and 20th angle subband of the transformed data shown in Fig.~\ref{fig:fff1}. It demonstrates that the curvelet coefficients vary continuously with respect to the spatial position $\mathbf{p}$. Furthermore, Fig.~\ref{fig:synchrosqueezing_reassignment} illustrates the distribution of curvelet coefficients across different scales and angles. Two distinct energy concentration regions associated with positive and negative angles are clearly visible, which perfectly corresponds to the original morphological characteristics of the crossing events in Fig.~\ref{fig:fff1}.

\multiplot*{2}{scoef_at_point,synchrosqueezing_reassignment}{width=0.45\columnwidth}
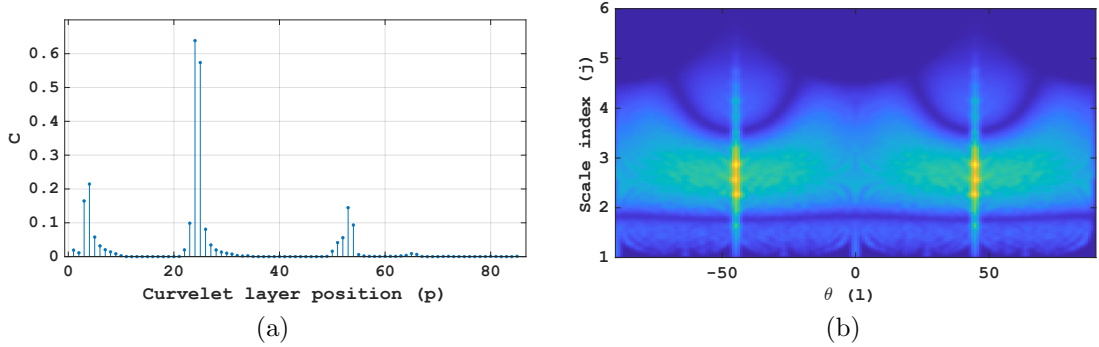
{Analysis of the curvelet transform results for Fig.~\ref{fig:fff1}: (a) coefficient amplitude variations with spatial position at a specific scale and angle subband; (b) energy distribution across different directions and scales.}

By applying the 2-D synchrosqueezing technique to map curvelet coefficients into the 4-D phase space ($\mathbf{p}-\boldsymbol{\omega}$ domain), it becomes possible to fully exploit the angular information inherent in the original seismic data. Signals originating from the same source display continuous angular variations, whereas interferences from blended sources exhibit discontinuities. Consequently, these signals can be effectively separated based on their distinct orientational characteristics within the curvelet domain. This procedure begins with the calculation of the 2-D instantaneous frequencies of the curvelet coefficients:
\begin{equation}
    \boldsymbol{\mu}(\mathbf{j}, \mathbf{l}, \mathbf{p}) = \operatorname{Real} \left\{ \frac{\nabla_{\mathbf{p}} \mathbf{C}(\mathbf{j}, \mathbf{l}, \mathbf{p})}{2\pi \mathrm{i} \, \mathbf{C}(\mathbf{j}, \mathbf{l}, \mathbf{p})}\right\}
\end{equation}
where $\nabla_\mathbf{p} \mathbf{C}(\mathbf{j}, \mathbf{l}, \mathbf{p})$ represents the gradient of the curvelet coefficients with respect to the spatial position $\mathbf{p}$, indicating the direction of steepest variation. The term $\boldsymbol{\mu}(\mathbf{j}, \mathbf{l}, \mathbf{p})$ denotes the resulting 2-D instantaneous frequency vector. By employing a squeezing operator, the energy $|\mathbf{C}(\mathbf{j}, \mathbf{l}, \mathbf{p})|^2$, which is initially smeared across the coarse parameter axes $(\mathbf{j}, \mathbf{l})$, is reallocated onto the precise physical frequency-wavenumber axes $\boldsymbol{\omega}$, yielding a multidimensional power spectrum:
\begin{equation}
    \mathbf{C}_{ls}(\boldsymbol{\omega}, \mathbf{p}) = \sum_{\mathbf{j}} \sum_{\mathbf{l}} |\mathbf{C}(\mathbf{j}, \mathbf{l}, \mathbf{p})|^2 \delta(\boldsymbol{\mu}(\mathbf{j}, \mathbf{l}, \mathbf{p}) - \boldsymbol{\omega})
\end{equation}
In this framework, while $\mathbf{C}_{ls}$ characterizes the energy density used for component identification, the corresponding complex curvelet coefficients are reassigned to these focal points to facilitate the reconstruction of deblended wavefields(Fig.~\ref{fig:synchrosqueezed_M_total}). 

\plot*{synchrosqueezed_M_total}{width=0.6\columnwidth}
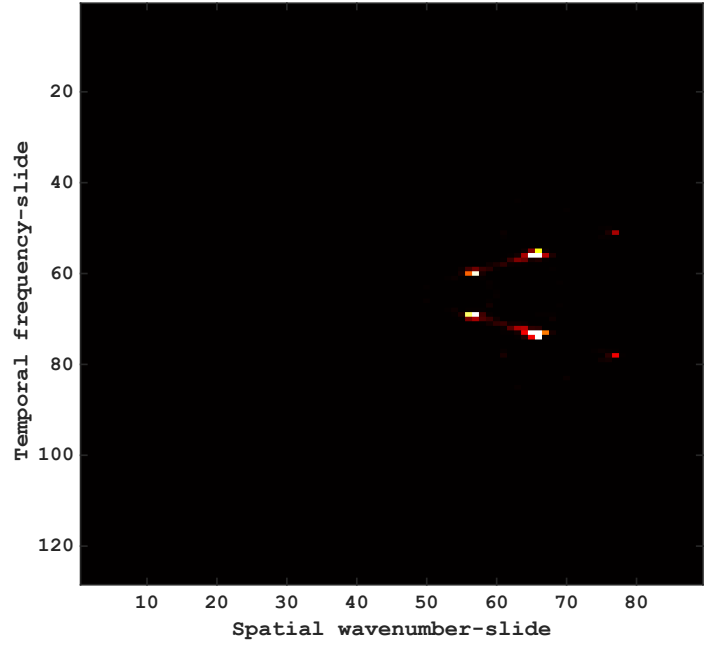
{The total synchrosqueezed energy spectrum, which corresponds to the spatial integration of $\mathbf{C}_{ls}(\boldsymbol{\omega}, \mathbf{p})$ across all spatial coordinates $\mathbf{p}$. Two distinct energy clusters, corresponding to the overlapping seismic events, can be clearly observed in the spectrum.}

Through this multi-dimensional phase space (MDPS) analysis, distinct energy clusters reflect kinematically coherent wave components sharing similar instantaneous frequencies. To accurately extract these components, a greedy energy-based clustering algorithm is implemented within the four-dimensional spectrum $\mathbf{C}_{ls}$. Initially, a predefined angular threshold is established to determine whether two points in the $f-k$ domain belong to the same coherent event. The algorithm then scans the spectrum to identify the point exhibiting the global maximum energy—representing the most prominent coherent wave component currently available—and designates it as the cluster ``center.'' All surrounding points satisfying the angular constraint relative to this center are assigned to this newly formed cluster. To prevent redundant grouping in subsequent steps, the energy of these clustered points is zeroed out (masked) within the working matrix. The algorithm then iteratively searches for the next local energy maximum in the residual matrix to form subsequent clusters. This iterative masking process continues until a predefined number of signal components is reached or the residual energy is completely depleted, marking the completion of the clustering stage. The intermediate maps generated from this clustering procedure are illustrated in Fig.~\ref{fig:num_group_map,global_clustering_kf}.

\multiplot*{2}{num_group_map,global_clustering_kf}{height=7cm}
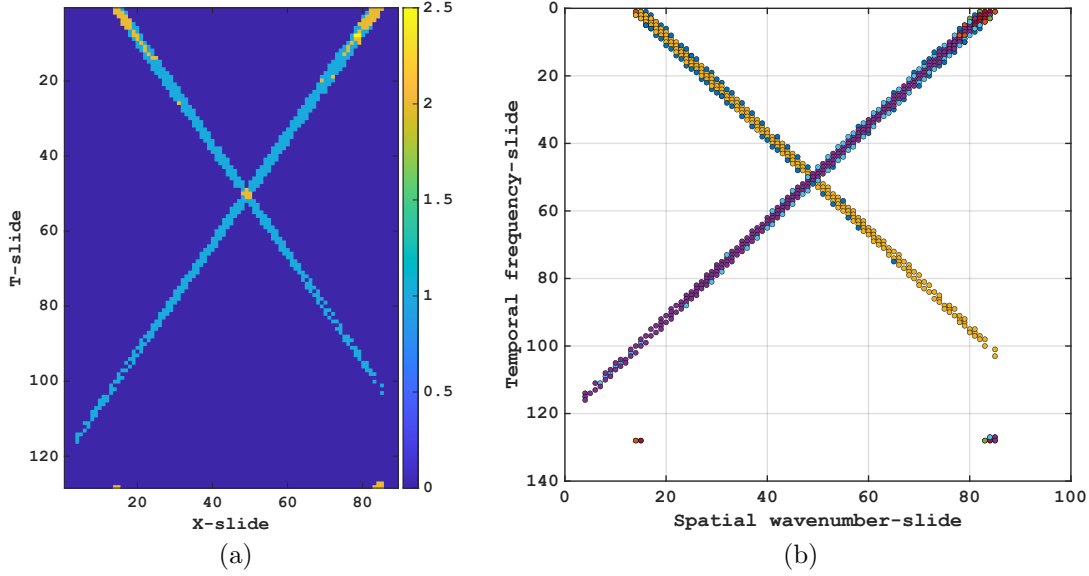
{Clustering results derived from the MDPS energy identification: (a) spatial distribution map of the identified energy components within local windows in the space-time domain, where higher values indicate a greater number of overlapping wave components; (b) distinct energy clusters partitioned within the reference $f-k$ domain, where each color represents a uniquely identified signal component.}

The extracted signal components derived from the clustering process are illustrated in Fig.~\ref{fig:extracted_coeff_energy_wave1,extracted_coeff_energy_wave2}. As clearly observed, the overlapping signals exhibiting distinct propagation directions have been successfully partitioned. In this specific example, the original curvelet coefficients were decomposed into ten distinct clusters. Subsequently, every five clusters sharing consistent directional characteristics were aggregated and assigned to represent a single source signal. Finally, by applying the inverse curvelet transform to these grouped coefficients, the separated individual source records are accurately recovered.

\multiplot*{2}{extracted_coeff_energy_wave1,extracted_coeff_energy_wave2}{height=5cm}
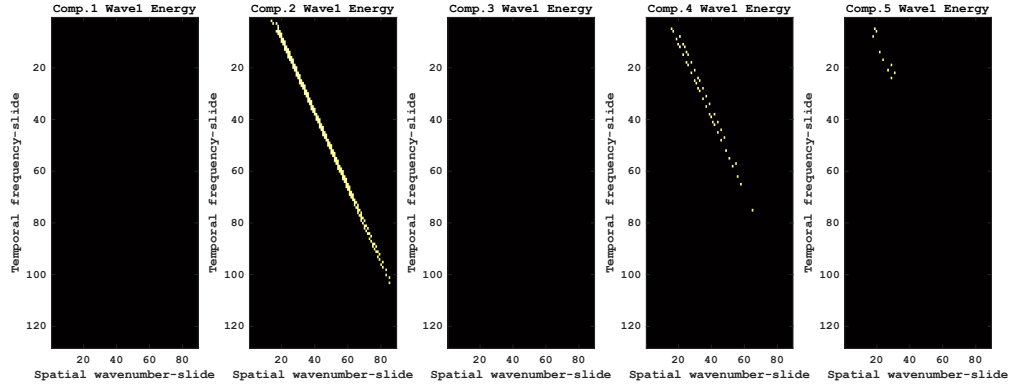
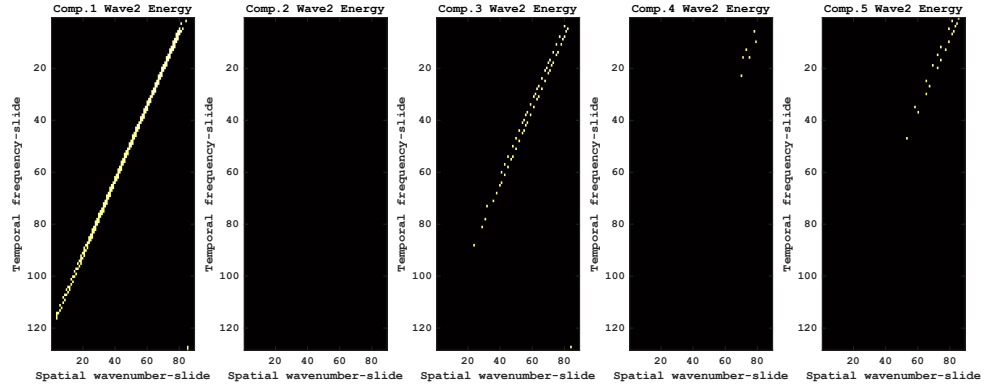
{Energy clusters allocated to the identified components: (a) grouped components corresponding to Source 1; (b) grouped components corresponding to Source 2.}

The separated data, which can be regarded as the direct output of the SSCT-constrained filtering, inevitably suffers from a certain degree of signal damage. Consequently, within the iterative framework, this filtered estimate is subsequently subtracted from the original raw data. This subtraction yields a pseudo-separated record that perfectly preserves the target signal without any structural damage, albeit still containing some residual blending noise, as illustrated in Fig.~\ref{fig:d1t}.

\subsection{Radon Transform}
During field acquisition, factors such as the absorption and attenuation of seismic waves often disrupt the continuity of reflected-wave events. These discontinuities pose significant challenges for the clustering process within the SSCT, potentially leading to signal degradation. Within the iterative deblending framework, the loss of target signal energy manifests as residual interference in the separated source records. Effectively attenuating this residual noise is crucial for further improving the separation quality. To achieve this, the Radon transform is introduced to further predict and remove the residual noise:
\begin{equation}
    u(\tau, p) = \int d(x, \tau + px) \, dx
\end{equation}
where $x$ denotes the spatial offset. In the $x-t$ domain, a seismic reflection event manifests as a coherent trajectory with a specific range of slownesses (slopes). By applying the linear Radon transform, various slowness values $p$ are systematically scanned to map the signal into the intercept-time and slowness domain ($\tau-p$ domain). When a specific $p$ value perfectly aligns with the true dip of a seismic event, the integration path of the transform operator tracks the event's trajectory exactly. Consequently, the constructive summation of amplitudes along this path produces highly focused energy maximums at the corresponding $(\tau, p)$ coordinates.

Achieving high-quality signal separation through this energy focusing requires the energy of the interference terms to be sufficiently small. Because the iterative SSCT successfully removes the bulk of the blending interference, the pseudo-separated records predominantly contain the complete primary signal, accompanied only by weak residual energy from the secondary source. By applying the Radon transform to these primary source records, the coherent primary signal focuses into compact energy clusters in the $\tau-p$ domain. In contrast, the residual secondary source energy remains unfocused and smeared. By applying a threshold or mask to isolate the focused primary energy, the cleaned $\tau-p$ coefficients are then projected back into the $x-t$ domain via the inverse Radon transform:
\begin{equation}
    \hat{d}(x, t) = \int \left[ u(\tau, p) * \rho(\tau) \right]_{\tau = t - px} \, dp
\end{equation}
where $\hat{d}(x, t)$ denotes the reconstructed, noise-free seismic record, and $\rho(\tau)$ represents the rho filter (a 1-D derivative-like operator) applied via convolution ($*$) to preserve correct signal amplitudes. By enforcing sparsity and focusing constraints in the Radon domain as a secondary separation stage within the iterative framework, a high-quality reconstruction of the coherent signals is ultimately achieved.

\multiplot*{3}{d1t,m,dm}{width=0.3\columnwidth}
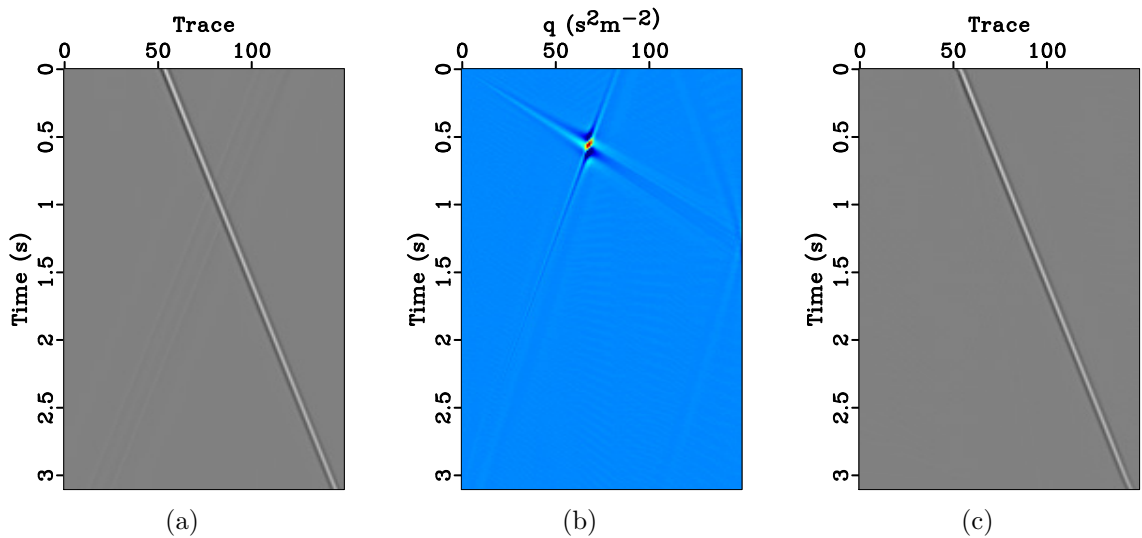
{Use Radon transform to predict and attenuate the residual noise.}

\section{Examples}
\subsection{Synthetic Examples}
To evaluate the effectiveness and robustness of the proposed method, a simple multi-layer geological model is utilized to simulate blended continuous recordings. Because source separation directly within the CSG domain is inherently challenging, the intermediate stages of the dual-constrained separation process are also presented to demonstrate the algorithm's performance. The unblended synthetic data (Fig.~\ref{fig:hybird}) and the corresponding blended data (Fig.~\ref{fig:blend}) used in this experiment are shown in Fig.~\ref{fig:hybird,blend}.

\multiplot*{3}{hybird,blend}{width=0.4\columnwidth}
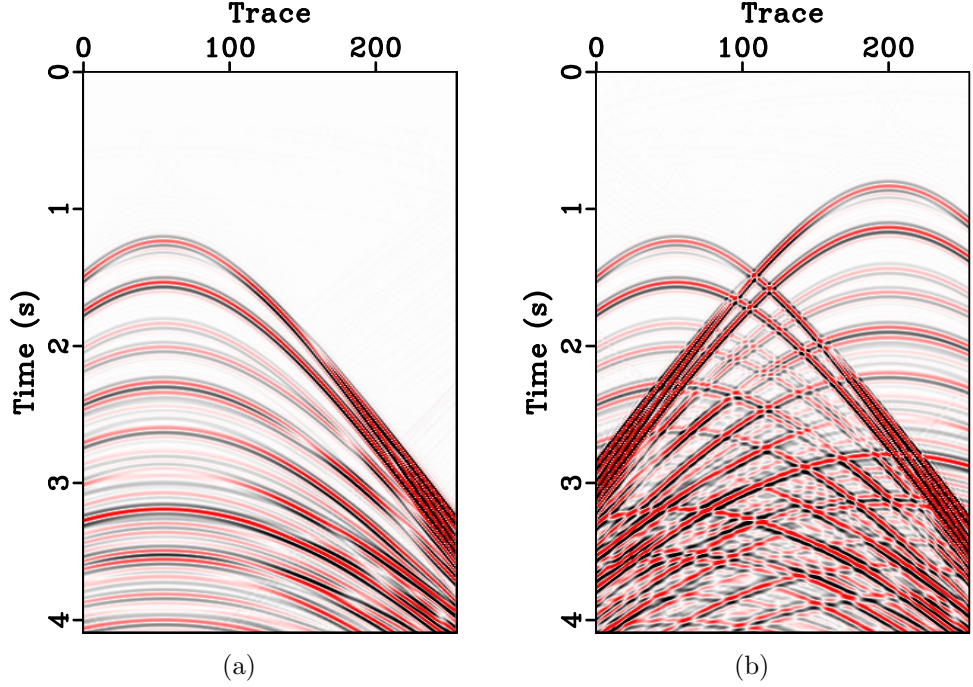
{Synthetic data used in the experiment: (a) unblended data; (b) blended data.}

The converged results of the iterative separation constrained by the SSCT are illustrated in Fig.~\ref{fig:deb_temp}, which displays the predicted deblended record after subtraction. As observed from the estimation error profile (Fig.~\ref{fig:loss_temp}), when the difference in dip angles between overlapping events is excessively small—particularly at the near offsets of the interfering source—the SSCT-based clustering process often struggles to accurately isolate the interference. This difficulty arises because near-offset reflections often exhibit significant curvature, reducing the sparsity of the signal in the transform domain and degrading the clustering accuracy. Furthermore, during the inverse transform following the energy synchrosqueezing, noticeable low-frequency residuals persist in the error profile. These low-frequency components are inherently challenging to decouple directly in the CSG domain due to the intrinsic construction of the curvelet transform itself. Specifically, at the coarsest scale (the lowest frequencies), curvelet basis functions are constructed as isotropic low-pass elements that fundamentally lack directional selectivity. Because the low-frequency components lose their distinct angular resolution within the curvelet frame, the SSCT-based clustering—which relies heavily on instantaneous orientational differences—fails to accurately distinguish the overlapping events, inevitably leading to the observed low-frequency signal leakage. Conversely, in higher-frequency regions exhibiting significant differences in dip angles, the SSCT achieves highly satisfactory separation results.

\multiplot*{2}{deb_temp,loss_temp}{width=0.4\columnwidth}
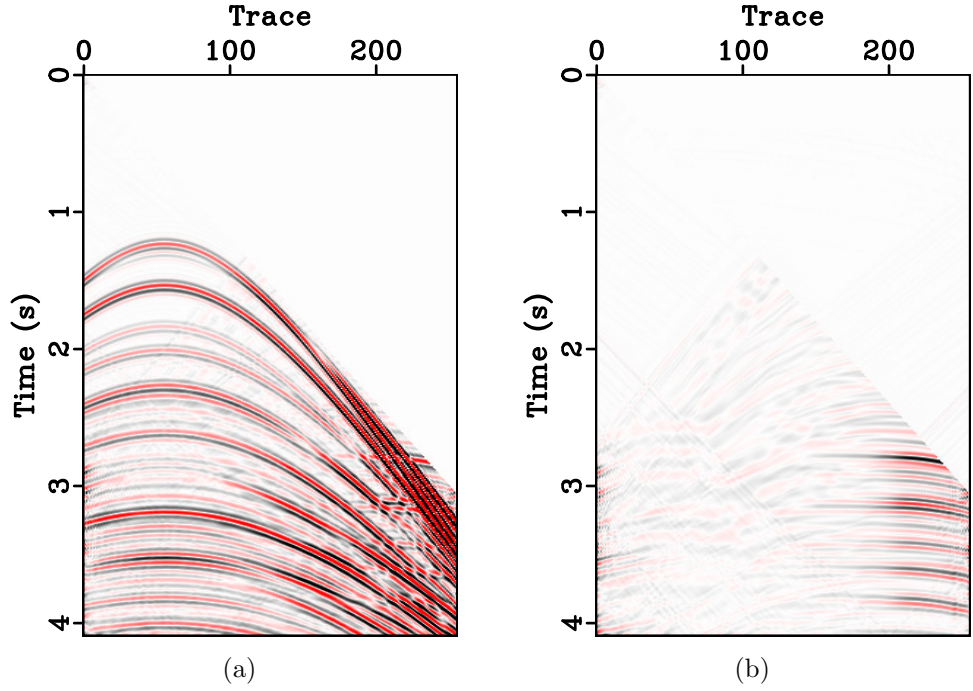
{Iterative separation results constrained by SSCT: (a) separated result; (b) estimation error.}

To further overcome the limitations of the SSCT, effective constraints in the Radon domain are subsequently employed to perform further iterative separation on the data once the SSCT iterations reach saturation (Fig.~\ref{fig:deb_temp}). Following the SSCT iterations, the residual interference in the severely contaminated regions can be distinctly distinguished from the coherent primary signals within the $\tau-p$ (intercept-time and slowness) domain. As clearly illustrated in Fig.~\ref{fig:radon_single1,radon_deb1,radon_diff}, when compared with the Radon spectrum of the clean data (Fig.~\ref{fig:radon_single1}), although the residual blending noise becomes resilient to further SSCT separation, the distinctly different $\tau-p$ ranges allow the focusing regions of the residual noise to be significantly separated from those of the effective signal in the Radon domain. Consequently, this residual noise is further attenuated, ultimately achieving high-fidelity deblending within the CSGs.

\multiplot*{3}{radon_single1,radon_deb1,radon_diff}{width=0.3\columnwidth}
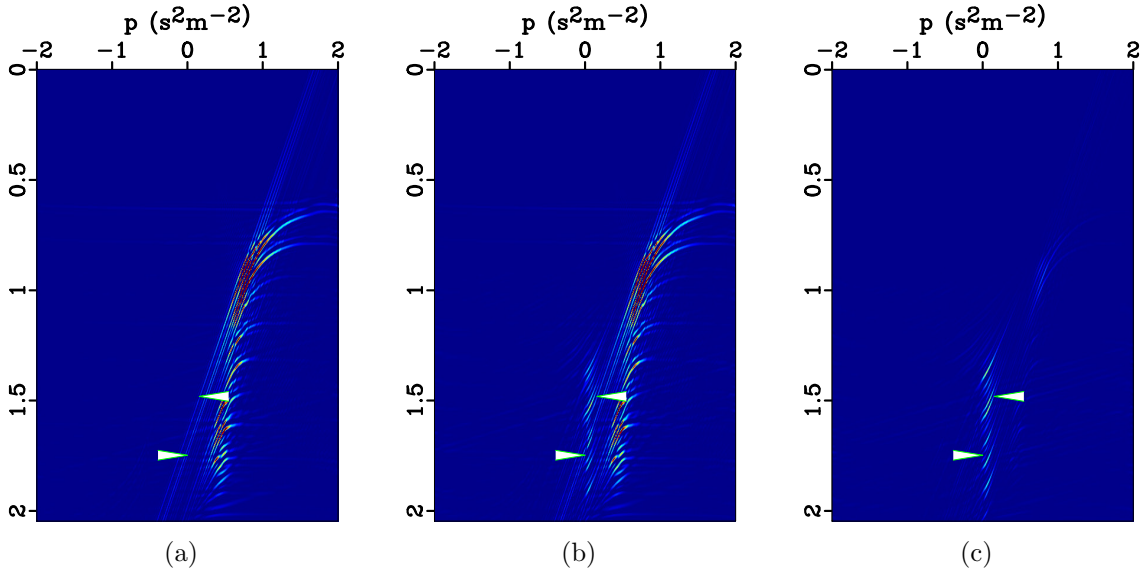
{Radon spectra of the corresponding data: (a) clean single-source data; (b) data after the SSCT iterations reach saturation; (c) Radon spectrum of the residual noise. It is evident that in the $\tau-p$ domain, the residual noise energy fails to focus constructively due to its distinct distribution range compared to the clean data.}

The final separation results are presented in Fig.~\ref{fig:hybird1,deb_final,loss1,hybird_fk2d,deb_final_fk2d,loss1_fk2d}. A direct comparison between Fig.~\ref{fig:deb_final} and Fig.~\ref{fig:radon_deb1} reveals that, following the subsequent iterations constrained within the Radon domain, the residual blending interference that previously resisted separation has been effectively attenuated. Furthermore, analysis in the frequency-wavenumber ($f-k$) domain confirms that signal leakage is restricted to a negligible level. The $f-k$ spectrum of the final separated result closely resembles that of the original unblended data, demonstrating that the proposed dual-constraint framework achieves high-fidelity deblending while successfully preserving the kinematic and frequency characteristics of the coherent signals.

\multiplot*{3}{hybird1,deb_final,loss1,hybird_fk2d,deb_final_fk2d,loss1_fk2d}{width=0.3\columnwidth}
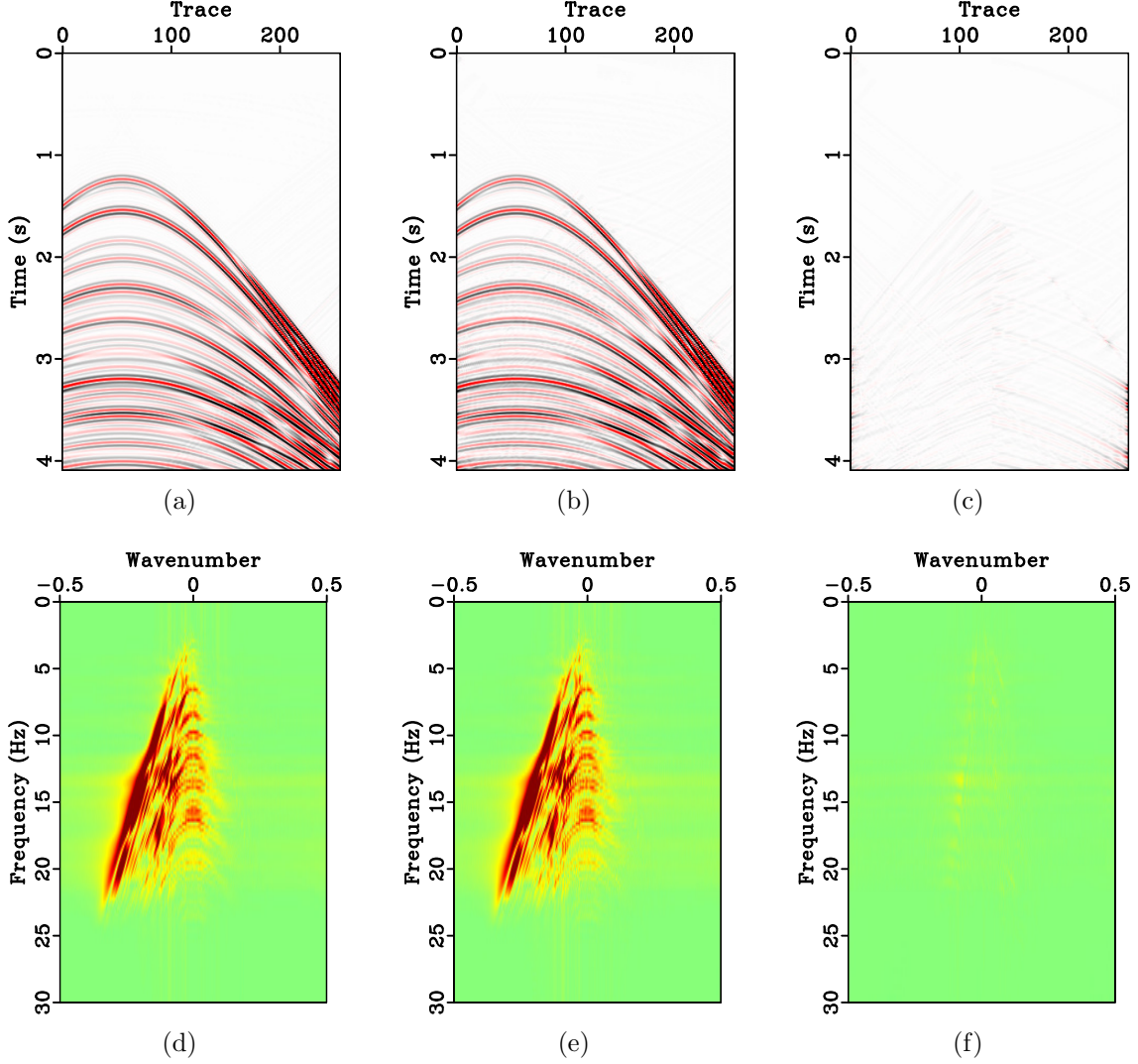
{Final deblending performance: (a) and (d) unblended data and its corresponding $f-k$ spectrum; (b) and (e) final separated result and its corresponding $f-k$ spectrum; (c) and (f) estimation error and its corresponding $f-k$ spectrum.}

\subsection{First Complex Data Experiment}
To evaluate the practical performance of the proposed algorithm, first test it on a real marine towed-streamer dataset that has been previously denoised and resampled. The data consists of 256 seismic traces, with 512 time samples per trace acquired at a 4 ms sampling interval. The original unblended data and the simulated blended data are displayed in Fig.~\ref{fig:single1_file1,blend_file1}. During the numerical blending simulation, the firing time delay between the two sources was deliberately set to zero. This zero-delay configuration maximizes the overlapping footprint of the interference within the common-shot gather (CSG), thereby providing a stringent worst-case scenario that thoroughly tests the separation capabilities of the proposed algorithm.

\multiplot*{2}{single1_file1,blend_file1}{width=0.45\columnwidth}
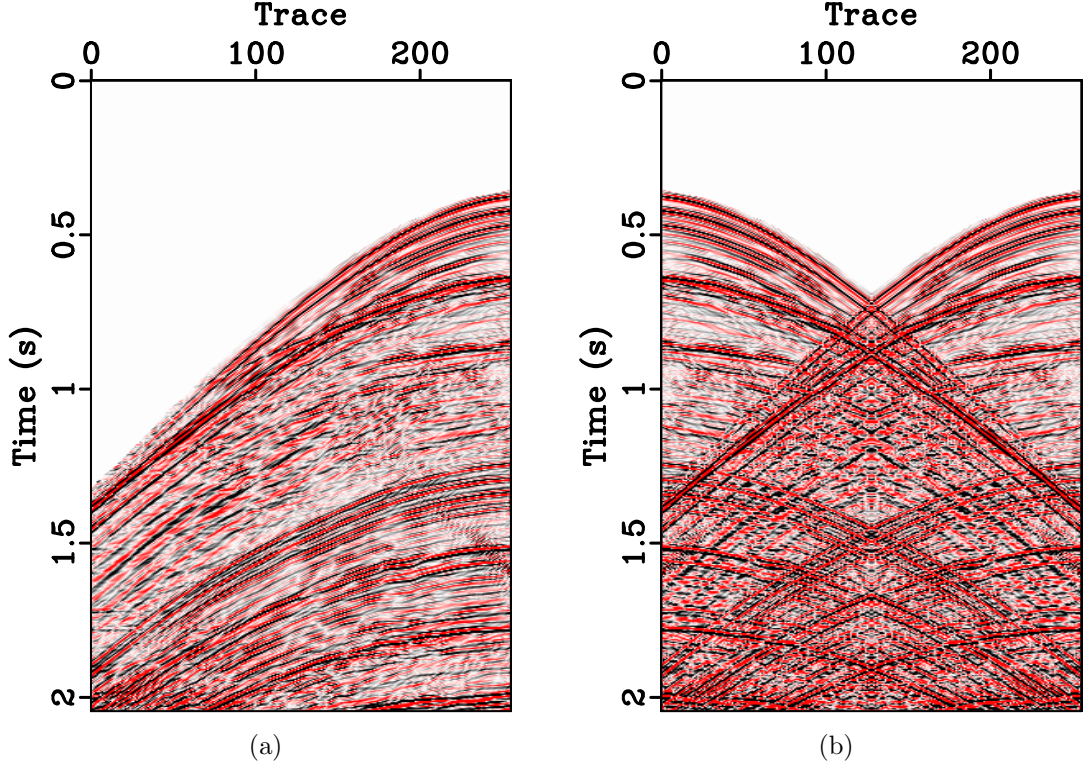
{The first marine field dataset used for testing: (a) unblended single-source data; (b) simulated blended data in the common-shot gather.}

The final separation results are presented in Fig.~\ref{fig:deb_final_file1}, achieving a signal-to-noise ratio (SNR) of 22.16 dB. As observed from the corresponding estimation error profile, the proposed algorithm performs exceptionally well on data characterized by clear, coherent reflection events. Signal leakage into the error section is tightly constrained and remains well within an acceptable range, validating the high-fidelity amplitude preservation of the dual-constraint approach.

\multiplot*{2}{deb_final_file1,loss_final_file1}{width=0.45\columnwidth}
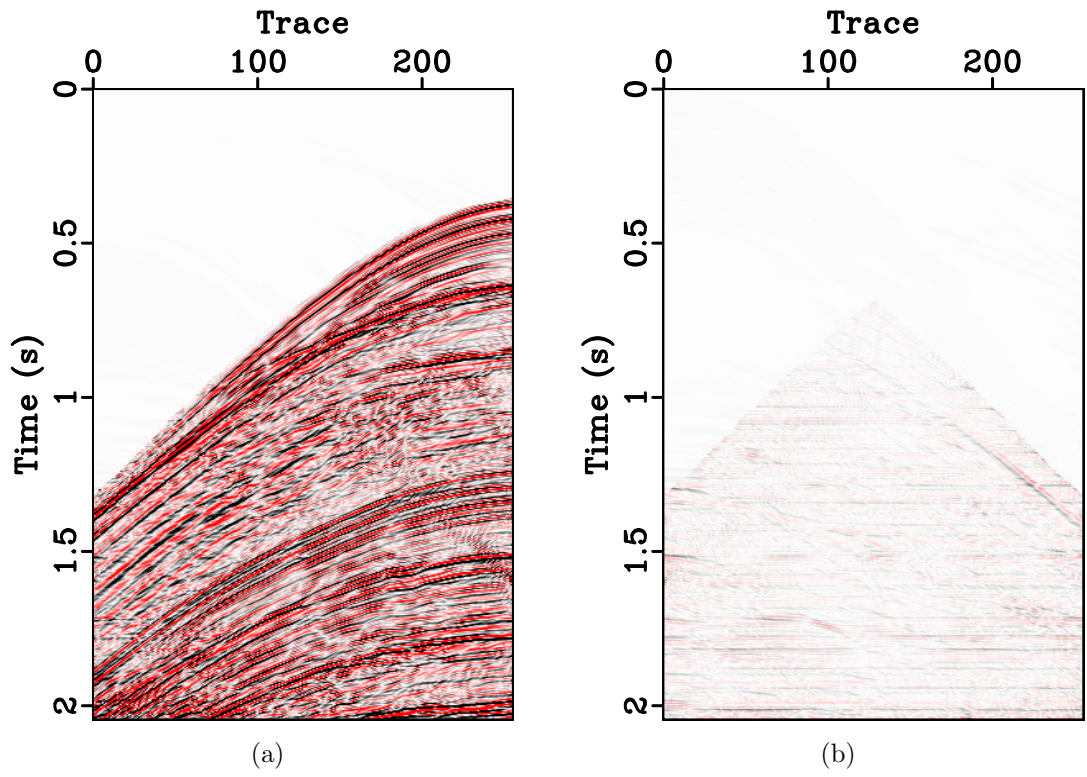
{Separation results using the proposed method on the field data: (a) deblended data; (b) estimation error.}

\subsection{Second Complex Data Experiment}
The second field dataset utilized in our experiments is a raw, unprocessed 3D seismic volume acquired via marine towed-streamer acquisition. It comprises 350 traces in the inline direction and 347 traces in the crossline direction, with 1120 time samples per trace recorded at an 8 ms sampling interval. For benchmarking purposes, the curvelet iterative shrinkage-thresholding algorithm(curvelet ISTA) was applied to the entire 3D dataset along the crossline direction (i.e., within the common-channel gather, CCG domain \cite[]{abmaSimultaneousSourceSeismic2020}). The final separation performance is subsequently evaluated and compared in the inline direction, specifically within the common-shot gather (CSG) domain.

The original unblended records and the simulated blended records are displayed in Fig.~\ref{fig:single1_file2,single_crg,blend_file2,blend_file2_crg_2}. During the numerical blending simulation, the range of the random firing time delays (time dithering) was deliberately restricted to a very narrow margin. This specific configuration severely challenges the curvelet ISTA. Traditional separation algorithms rely heavily on large random time dithers to break the coherency of the blending interference. When the variations in the delay times are exceptionally small, they often fail to exceed the threshold required to render the interference sufficiently incoherent (a constraint closely related to Jiang's limit), which inevitably leads to compromised separation accuracy. The specific time-dithering curve applied to the sources is plotted in Fig.~\ref{fig:blend_file2_crg_2}.

\multiplot*{2}{single1_file2,single_crg,blend_file2,blend_file2_crg_2}{width=0.3\columnwidth}
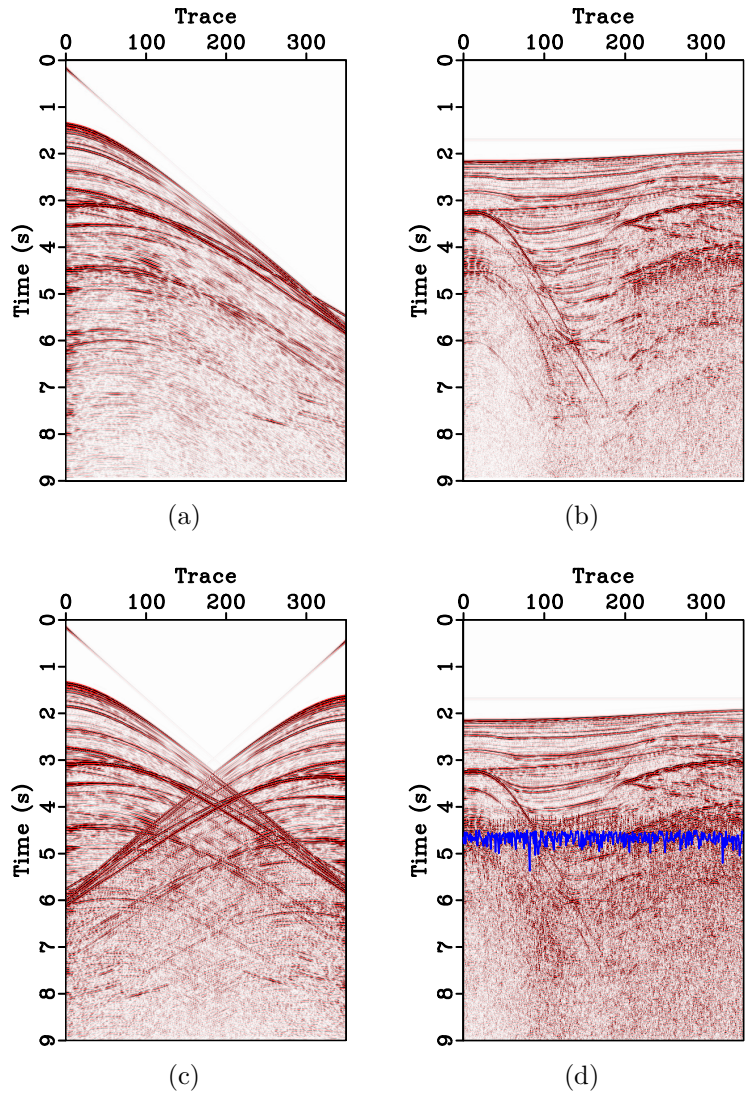
{The second field dataset and its corresponding blended records: (a) unblended CSG record; (b) unblended CCG record; (c) blended CSG record; (d) blended CCG record.}

A detailed comparison of the separation performance between the  the curvelet ISTA and the proposed method is illustrated in Fig.~\ref{fig:deb_cur_file2-window,deb_final_file2-window,deb_cur_file2-loss-window,deb_final_file2-loss-window,deb_cur_file2_fk2d,deb_final_file2_fk2d,deb_cur_file2-loss_fk2d,deb_final_file2-loss_fk2d}. As observed, the inability to sufficiently break the coherency of the blending interference—caused by the deliberately restricted small time dithers—visibly compromises the separation results of the curvelet ISTA. The comparison of the $f-k$ spectrum estimation errors (Fig.~\ref{fig:deb_cur_file2-loss_fk2d} and \ref{fig:deb_final_file2-loss_fk2d}) further reveals that when the inherent continuity of the data is poor and the time-dithering margin is excessively narrow, the curvelet ISTA separation suffers from severe signal leakage.

This degradation is particularly pronounced in severely blended regions where direct waves or primary reflections strongly intersect. As highlighted in the local zoom-in sections displayed in Fig.~\ref{fig:deb_cur_file2-local,deb_final_file2-local,deb_cur_file2-loss-local,deb_final_file2-loss-local}, the separation quality of the curvelet ISTA method is highly unsatisfactory in these complex crossing areas. These localized regions in the CSG generally correspond to the areas with the maximum overlapping footprint of the two sources in the CCG domain. Such extensive blending interference, compounded by the extremely low coherency differences (due to inadequate random time delays), fundamentally degrades the separation fidelity of the curvelet ISTA. In contrast, the proposed method successfully preserves the signal integrity.

\multiplot*{4}{deb_cur_file2-window,deb_final_file2-window,deb_cur_file2-loss-window,deb_final_file2-loss-window,deb_cur_file2_fk2d,deb_final_file2_fk2d,deb_cur_file2-loss_fk2d,deb_final_file2-loss_fk2d}{width=0.2\columnwidth}
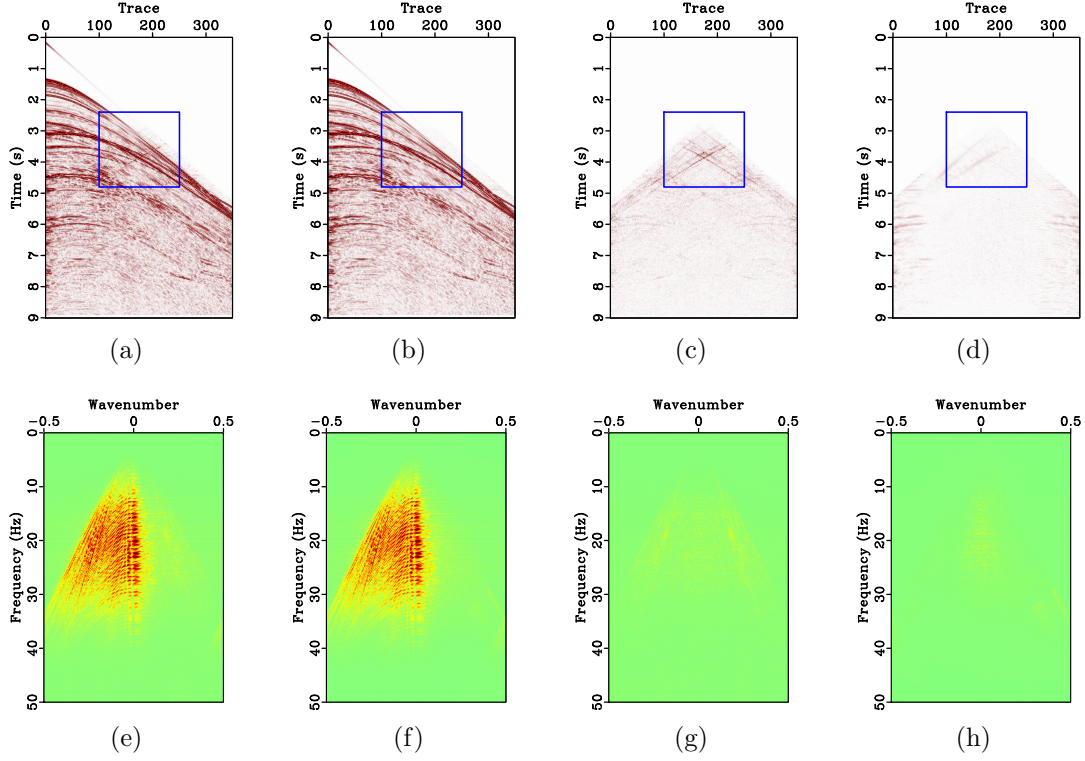
{Comparison of the separation performance in the CSG domain. Results obtained using the conventional curvelet ISTA: (a) separated data with an SNR of 12.95 dB, (c) estimation error, (e) corresponding $f-k$ amplitude spectrum, and (g) $f-k$ spectrum of the error. Results achieved by the proposed method: (b) separated data with an SNR of 23.93 dB, (d) estimation error, (f) corresponding $f-k$ amplitude spectrum, and (h) $f-k$ spectrum of the error.}

\multiplot*{2}{deb_cur_file2-local,deb_final_file2-local,deb_cur_file2-loss-local,deb_final_file2-loss-local}{width=0.4\columnwidth}
{Local zoom-in sections from: (a) Fig.~\ref{fig:deb_cur_file2-window}; (b) Fig.~\ref{fig:deb_final_file2-window}; (c) Fig.~\ref{fig:deb_cur_file2-loss-window}; (d) Fig.~\ref{fig:deb_final_file2-loss-window}.}

Finally, to conclude the experiments, the separation results in the common-channel gather (CCG) domain obtained by the iterative curvelet thresholding and the proposed method are compared. The Fig.~\ref{fig:blend_file2_crg_2} displays the 173 $rd$ CCG profile, which represents the region suffering from the most extensive blending overlap and the most severe direct-wave interference. The corresponding clean and blended data are illustrated in Fig.~\ref{fig:single_crg} and Fig.~\ref{fig:blend_file2_crg_2}, respectively. As observed in Figs.~\ref{fig:deb_cur_crg}, \ref{fig:deb_syn_crg}, \ref{fig:deb_cur_crg-loss}, and \ref{fig:deb_syn_crg-loss}, the SNR of the separation result achieved by the conventional iterative curvelet thresholding on the CCG data is 14.65 dB, whereas the SNR yielded by the proposed method reaches 20.21 dB. Conventional algorithms, which rely solely on the coherence differences introduced by time dithering, often struggle to reconstruct fragmented seismic events and handle intense blending interference within the profile. In contrast, the proposed CSG-based iterative separation framework demonstrates a significantly superior capability in preserving these weak and fragmented signals from the original unblended data.

\multiplot*{2}{deb_cur_crg,deb_syn_crg,deb_cur_crg-loss,deb_syn_crg-loss}{width=0.35\columnwidth}
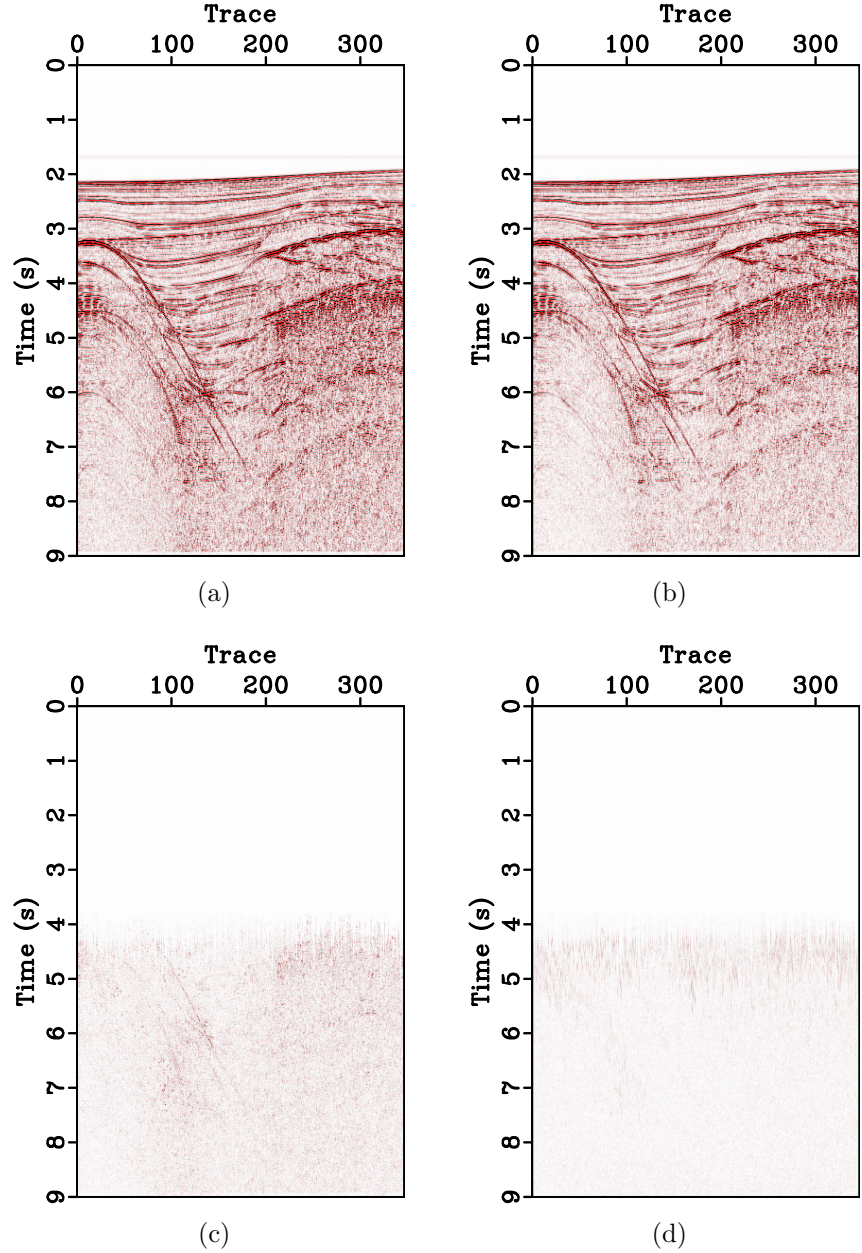
{Comparison of the separation performance in the CCG domain. Conventional curvelet transform results: (a) separated data, with an SNR of 14.65 dB; (c) estimation error; Proposed method results: (b) separated data, with a SNR of 20.21 dB;(d) estimation error.}

\section{Conclusion}
In this study, we proposed a novel dual-constrained iterative framework to tackle the inherently challenging problem of separating blended seismic data directly within the common-shot gather (CSG) domain. Conventional separation methods, such as standard iterative curvelet thresholding, rely heavily on large random firing time delays (time dithering) to sufficiently break the coherency of the blending interference, a requirement rigorously governed by Jiang's limit. When these time delays are restricted to a very narrow margin, or when processing low-frequency components that intrinsically require much larger time shifts to become completely incoherent, traditional methods suffer from severe signal leakage and compromised amplitude fidelity.

To overcome these fundamental limitations, our proposed framework seamlessly integrates the synchrosqueezed curvelet transform (SSCT) and the Radon transform into a direct-subtraction iterative process. Initially, the SSCT projects the continuous seismic records into a multi-dimensional phase space (MDPS). By systematically evaluating localized spatial distances and instantaneous angular differences, the SSCT-based clustering robustly isolates the coherent target signals from the interference. Furthermore, to rigorously mitigate the intrinsic geometric limitations of the curvelet transform at the coarsest scales—where low-frequency elements act as isotropic low-pass filters lacking directional selectivity—we introduced a secondary filtering stage in the intercept-slowness ($\tau-p$) domain using the Radon transform. Because the residual blending noise fails to focus constructively in the $\tau-p$ domain due to its distinct kinematic characteristics, it is effectively attenuated via Radon thresholding, leaving the primary coherent signal completely intact.

Rigorous evaluations using both synthetic models and complex marine towed-streamer field datasets validated the superiority of the proposed approach. Even in extreme, worst-case scenarios—such as zero-delay firing configurations or highly restricted time dithers that typically cause traditional curvelet methods to fail at complex crossing events—the proposed SSCT-Radon joint framework successfully achieved high-fidelity source separation. The estimation error profiles and corresponding frequency-wavenumber ($f-k$) spectra confirmed that our method significantly minimizes signal leakage and flawlessly preserves both the kinematic and amplitude characteristics of the data. Ultimately, this dual-constrained strategy largely liberates the deblending process from the stringent dependency on large random delay-time differences, offering a highly robust and pristine reconstruction solution for modern simultaneous-source seismic acquisition.

\bibliographystyle{seg}  
\bibliography{example}

\end{document}